\documentclass[article]{aa}
\usepackage{natbib}
\usepackage{graphicx}
\usepackage{rotating}
\usepackage{times}
\usepackage{amsmath}
\usepackage{txfonts}
\usepackage{multirow}
\newcommand{\nht}{\ifmmode {{\rm NH}_3} \else {NH{\bas 3}} \fi}
\newcommand{\tco}{\ifmmode {^{13}{\rm CO}} \else {$^{13}{\rm CO}$}\fi}
\newcommand{\dco}{\ifmmode {^{12}{\rm CO}} \else {$^{12}{\rm CO}$}\fi}
\newcommand{\cdo}{\ifmmode {{\rm C}^{18}{\rm O}} \else {${\rm C}^{18}{\rm O}$}\fi}

\newcommand{\htco}{\ifmmode {{\rm H}^{13}{\rm CO}^{+} } \else {${\rm H}^{13}
{\rm CO}^{+}$ }\fi}
\newcommand{\hco}{\ifmmode {{\rm H}^{12}{\rm CO}^{+} } \else {${\rm H}^{12}
{\rm CO}^{+}$ }\fi}
\newcommand{\juz}{\ifmmode {{\rm J}=1\rightarrow 0} \else
{J=1$\rightarrow$0}\fi}
\newcommand{\jdu}{\ifmmode {{\rm J}=2\rightarrow 1} \else
{J=2$\rightarrow$1}\fi}
\newcommand{\jtd}{\ifmmode {{\rm J}=3\rightarrow 2} \else
{J=3$\rightarrow$2} \fi}
\newcommand{\jcq}{\ifmmode {{\rm J}=5\!\rightarrow\!4} \else
{${\rm J}=5\!\rightarrow\!4$} \fi}
\newcommand{\as}{\ifmmode {^{\scriptscriptstyle\prime\prime}}
        \else $^{\scriptscriptstyle\prime\prime}$\fi}
\newcommand{\am}{\ifmmode {^{\scriptscriptstyle\prime}}
        \else $^{\scriptscriptstyle\prime}$\fi}
\newcommand{\hh}{\ifmmode {{\rm H}_2} \else {H$_2$} \fi}
\renewcommand{\hco}{\ifmmode {{\rm HCO}^+} \else {HCO$^+$} \fi}
\newcommand{\hhco}{\ifmmode {{\rm H}_2{\rm CO}} \else {H$_2$CO} \fi}
\newcommand{\ddco}{\ifmmode {{\rm D}_2{\rm CO}} \else {D$_2$CO} \fi}
\newcommand{\chhdoh}{\ifmmode {{\rm CH}_2{\rm DOH}^+} \else {CH$_2$DOH} \fi}
\newcommand{\chhhod}{\ifmmode {{\rm CH}_3{\rm OD}^+} \else {CH$_3$OD} \fi}
\newcommand{\chhhoh}{\ifmmode {{\rm CH}_3{\rm OH}^+} \else {CH$_3$OH} \fi}
\newcommand{\tchhhoh}{\ifmmode {^{13}{\rm CH}_3{\rm OH}^+} \else {$^{13}$CH$_3$OH} \fi}
\newcommand{\dcop}{\ifmmode {{\rm DCO}^+} \else {DCO$^+$} \fi}

\newcommand{\hhd}{H$_2$D$^+$}
\newcommand{\ohhd}{o-H$_2$D$^+$}
\renewcommand{\textbf}{\textrm}
\begin{document}
\title{A deep search for H$_2$D$^+$ in protoplanetary disks
\thanks{Based on observations carried out with the Atacama Pathfinder Experiment and the James Clerk Maxwell Telescope. APEX is a collaboration between the Max-Planck-Institut f\"ur Radioastronomie, the European Southern Observatory, and the Onsala Space Observatory. The JCMT is operated by the Joint Astronomy Centre on behalf of the Science and Technology Facilities Council of the United Kingdom, the Netherlands Organisation for Scientific Research, and the National Research Council of Canada. }
}
\subtitle{Perspectives for ALMA}

\author{
Edwige Chapillon\inst{1}\thanks{\emph{Present address:}
Institute of Astronomy and Astrophysics, Academia Sinica, P.O. Box 23-141, Taipei 106, Taiwan, ROC, \email{chapillon@asiaa.sinica.edu.tw}}
\and B\'ereng\`ere Parise\inst{1}
\and St\'ephane Guilloteau \inst{2,3}
\and Fujun Du\inst{1}
}
%
%offprints{E.Chapillon \email{chapillon@asiaa.sinica.edu.tw}}
%
\institute{ 
 MPIfR, Auf dem H\"ugel 69, 53121 Bonn, Germany.\\
 \email{echapill@mpifr-bonn.mpg.de, bparise@mpifr-bonn.mpg.de, fjdu@mpifr-bonn.mpg.de}
\and{}
Universit\'e de Bordeaux, Observatoire Aquitain des Sciences de l'Univers, 2 rue de l'Observatoire BP 89, F-33271 Floirac, France
 \and{}
CNRS/INSU - UMR5804, Laboratoire d'Astrophysique de Bordeaux;  2 rue de l'Observatoire BP 89, F-33271 Floirac, France \\
  \email{guilloteau@obs.u-bordeaux1.fr}
%Institute of Astronomy and Astrophysics, Academia Sinica, P.O. Box 23-141, Taipei 106, Taiwan, ROC\\
%\email{chapillon@asiaa.sinica.edu.tw}
}

\date{Received **-***-****, Accepted **-***-****}
\authorrunning{}
\titlerunning{}

  \abstract
% context heading (optional)
% {} leave it empty if  necessary
{The structure in density and temperature of protoplanetary disks surrounding low-mass stars is not yet well known. 
The protoplanetary disks mid-planes are expected to be very cold and thus depleted in molecules in gas phase, especially CO. Recent observations of molecules at very low apparent temperature ($\sim 6$\,K) challenge this current picture of the protoplanetary disk structures. 
}
% aims heading (mandatory)
{We aim at constraining the physical conditions, and in particular the gas-phase CO abundance in the mid-plane of protoplanetary disks.
}
% methods heading (mandatory)
{The light molecule \hhd\,is a tracer of cold and CO-depleted environment. It is therefore a good candidate to explore the disks mid-planes. We performed a deep search for \hhd\,in the two well-known disks surrounding TW\,Hya and DM\,Tau using the APEX and JCMT telescopes. The analysis of the observations are done with DISKFIT, a radiative transfer code dedicated to disks. 
In addition, we used a chemical model describing deuterium chemistry to infer the implications of our observations on the level of CO depletion and on the ionization rate in the disks mid-plane.
%An additional chemical modeling is done {to constrain the physical conditions, and in particular ionization and CO abundance in the disks mid-plane.} 
}
% results heading (mandatory)
{The ortho-\hhd\,$(1_{1,0}-1_{1,1})$ line at 372\,GHz was not detected. Although our limit is three times better than previous observations, comparison with the chemical modeling indicates that it remains insufficient to put valuable constraints on the CO abundance in the disk mid-plane. 
%\hhd\,will be extremely difficult to detect even with a powerful instrument such as ALMA.
}
% conclusions heading (optional), leave it empty if necessary
{{Even with ALMA, the detection of \hhd\ may not be straightforward, and \hhd\ may not be a sufficiently sensitive tracer of the protoplanetary disks mid-plane.}}


%----------------------------------
\maketitle{}
%----------------------------------

%-------------------------------------------------------------------
\section{Introduction}
Planetary systems are believed to be formed in the latest stages of protostellar evolution, within protoplanetary disks. The details of this process are however not known. Constraining the physical and chemical structure in  disks during their evolution are requirements to get more insights into this question.  

Current models predict that protoplanetary disks consist of three typical layers, which are, going from the surface to the mid-plane of the disk: 1) the outer layer, a PDR directly illuminated by the stellar UV, 2) a warm molecular zone, where the UV is sufficiently attenuated to allow formation of molecules, and 3) a cold mid-plane where temperature is low and most molecules are expected to be frozen on dust grains \citep{Bergin_etal2007}. 
All static chemical models published so far produce similar results \citep[e.g.][]{Aikawa_Nomura_2006, Semenov_etal2005}. The chemistry is dominated by two main processes: molecular freeze-out onto grains in the cold mid-plane and photodissociation in the upper layers (enhanced by grain growth that allows the UV field to penetrate deeper in the disk). 
This simple picture is however not yet constrained by observations, and some puzzles still remain. In particular, very cold C$_2$H, CN and CO ($\leq$15\,K, i.e. below the evaporation temperature) were detected in the DM\,Tau disk \citep{Dartois_etal2003, Pietu_etal2007, Henning_etal2010, Chapillon_etal2011}, pointing out that some process may lead to cold gas-phase molecules in the disk mid-plane. 
Radial and vertical mixing is usually invoked to explain partial molecular replenishment of the cold disk mid-plane from the higher warm layer \citep{Semenov_etal2006, Aikawa_2007}. Such an explanation is not fully satisfactory with respect to the current chemical results \citep[see for example][]{Hersant_etal2009}. Observations of a chemical tracer of cold regions are thus needed.%%\\

\hhd\,is a very promising molecule to trace  the disk mid-plane as it is likely the only remaining observable species in cold medium, where molecules are frozen onto grains.  H$_2$D$^+$ is exclusively  formed in the gas phase at low temperatures (T$<$\,20--30K). Moreover, CO (and other heavy species like N$_2$) is a very efficient destroyer of H$_3^+$ and \hhd. High abundances of \hhd\,can thus only build up in depleted environments. 
The ortho-\hhd\,$(1_{1,0}-1_{1,1})$ line, hereafter \ohhd, is commonly detected towards cold and dense pre-stellar cores, see for instance \citet{Caselli_etal2008}. 
Observation of \ohhd\,was also proposed by \citet{Ceccarelli_etal2004} as a way to measure the mid-plane ionization, since \hhd\,may %should 
be the dominant ion in that region. %{\it ou plutot H3+ dominant, mais pas observable (ou D3+ d'apres Willacy 2007??)}.

Until now, no firm detection of \ohhd\,in protoplanetary disks has been definitively secured. A tentative detection on DM\,Tau obtained with the CSO was claimed by \citet{Ceccarelli_etal2004} and later challenged by SMA observations by \citet{Qi_etal2008}. Using a line profile appropriate for the DM Tau disk, \citet{Guilloteau_etal2006} showed that the \citet{Ceccarelli_etal2004} data indicates only a $2 \sigma$ tentative detection. There are only upper-limits in TW\,Hya \citep{Ceccarelli_etal2004,Thi_etal2004,Qi_etal2008} obtained with the JCMT and the SMA. 
%The observations are presented in Sect.\,\ref{sec:obs}, and analysed in Sect.\,\ref{sec:res}. Additional chemical modeling is presented in Sect.\,\ref{sec:dis}, the feasibility of ALMA observations are discussed in Sect.\,\ref{sec:alma} and we summarize our study in Sect.\,\ref{sec:cc}

The paper is organized as follows: the observations are presented in Sect.\,\ref{sec:obs}, and upper limits on the \ohhd\,column densities are derived in Sect.\,\ref{sec:res}. The implications %of the observational upper limits 
on the level of CO depletion and ionization in the disk midplane are discussed 
%in the light of a chemical model 
in Sect.\,\ref{sec:dis}. 
The feasibility of ALMA observations are discussed in Sect.\,\ref{sec:alma} and we summarize our study in Sect.\,\ref{sec:cc}.

\section{Observations}
\label{sec:obs}
%\subsection{Sources}
We searched for \ohhd\,in two well-studied, close or large protoplanetary disks. Their properties are summarized in Table \ref{tab:source}.

 TW\,Hya is the  nearest observable T-Tauri star, located only at 56\,pc from the Sun, and surrounded by a rather small disk of about 200\,AU of radius. It is the only protoplanetary disk where molecules have been studied in some systematic way because of its proximity \citep[][]{Qi_etal2008}. The disk is viewed nearly face-on. 

 DM\,Tau is a well known TTauri star of 0.5 $M_\odot$ surrounded by a large and massive molecular disk \citep{Guilloteau_Dutrey_1994,Guilloteau_Dutrey_1998, Oberg_etal2010}.
\citet{Dartois_etal2003} and \citet{Pietu_etal2007} have performed  high resolution multi-transition, multi-isotope studies of CO, and demonstrated the existence of a vertical temperature gradient in the disk as well as the presence of cold gas-phase CO. %%\\

\begin{table*}
\caption{Sources coordinates (J2000) and properties}
\small
\begin{tabular}{lcccccccc}
\hline \hline
Source & RA & DEC & V$_{lsr}$ (km\,s$^{-1}$) & Distance (pc) & Radius\tablefootmark{*} (AU) &$\Delta_v$\tablefootmark{**} (km\,s$^{-1}$) & M$_*$ (M$_\odot$) & M$_{disk} $ ($10^{-2}$\,M$_\odot$)\tablefootmark{***}  \\
\hline
DM\,Tau & 04:33:48.733 &  18:10:09.890  & 6.1 & 140 & 800 & 2 &0.5&3 \\
TW\,Hya & 11:01:51.875 & -34:42:17.155 & 2.7 & 56 & 200 & 1 &0.5& 3 \\
\hline
\end{tabular}
\normalsize
\tablefoot{
\tablefoottext{*}{From CO observations. }
\tablefoottext{**}{FWHM derived from CO observations}
\tablefoottext{***}{From continuum observations  \citep{Dutrey_etal1997,Wilner_etal2000} assuming a dust mass absorption coeffficient $\kappa(\nu) = 0.1\times(\nu/10^{12}\mathrm{Hz})^{-1}$ cm$^2$/g. The uncertainties are at least a factor $\pm3$ due to the unknown dust characteristics.
}
}
\label{tab:source}
\end{table*}

%\subsection{DM\,Tau --- JCMT data}
%\paragraph{DM\,Tau --- JCMT data:}
We retrieved observations of \ohhd\,toward DM\,Tau from the JCMT archive. 
%DM\,Tau was observed with the JCMT telescope. Data are now public.
The \ohhd\,line at 372.421\,GHz was observed together with L1544 in January 2007 with the multi-beam receiver HARP and the ACSIS backend with a resolution of 0.03\,km\,s$^{-1}$. The source was centered in the pixel at the tracking center (H10). DM\,Tau was observed during about 10.75\,h on source, with {a} system temperature of 605\,K on average leading to a rms of $44$\,mK (T$_A^*$) at {a resolution of }0.03\,km\,s$^{-1}$.

%\begin{figure}
%\centering
%\includegraphics[angle=270,width=6cm]{h2dp-dmtau.eps}\\
%\includegraphics[angle=270,width=6cm]{h2dp-twhya.eps}
%\caption{
%\hhd\,spectrum in DM\,Tau.  The resolution is 0.17\,km\,s$^{-1}$
%\hhd\,spectrum in TW\,Hya.The resolution is 0.20\,km\,s$^{-1}$.
% The dashed line indicate the systemic velocity. 
%}
%\end{figure}

\begin{figure}
\centering
{\includegraphics[angle=270,width=7cm]{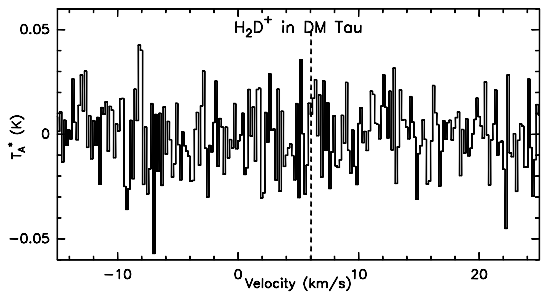}}
\caption{\ohhd\,spectrum in DM\,Tau. The dashed line indicate the systemic velocity. The resolution is 0.17\,km\,s$^{-1}$}
\label{fig:obs-dmtau}
\end{figure}

%\subsection{TW\,Hya --- APEX data}
%\paragraph{TW\,Hya --- APEX data:}
Using the APEX telescope, we observed the \ohhd\, at 372.421\,GHz toward TW\,Hya during 3.8\,h on source in July and September 2010 with the new FLASH\,345 receiver \citep[the dual-polarization receiver FLASH operates a two SB SIS mixer provided by IRAM, see][]{Maier_etal2005}.
%The receiver was tuned at 372.42\,GHz in USB mode. 
The July and September observations were carried out with different backends, leading respectively to a spectral resolution of 0.15\,km\,s$^{-1}$ and 0.06\,km\,s$^{-1}$. Data were resampled to a spectral resolution of 0.20\,km\,s$^{-1}$. As no planet was visible, the focus was checked by line pointing on RAFGL\,5254 and  IRAS\,07454-7112, pointing was done toward the same calibrators. The rms noise is 18\,mK(T$_A^*$) obtained after removing a constant baseline to the data at 0.2\,km\,s$^{-1}$ of resolution.%%\\

\begin{figure}
\centering
{\includegraphics[angle=270,width=7cm]{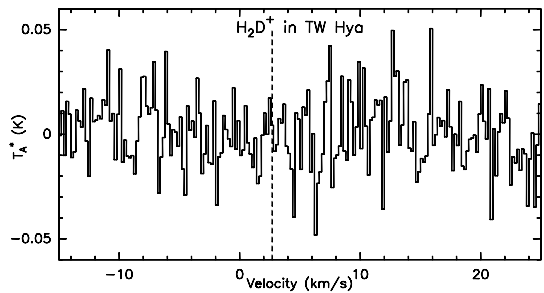}}
\caption{\ohhd\,spectrum in TW\,Hya. The dashed line indicate the systemic velocity. The resolution is 0.20\,km\,s$^{-1}$.}
\label{fig:obs-twhya}
\end{figure}

%\subsection{Results}
%No detection was performed in any of the sources. 
{We did not detect any \ohhd\ emission in any of the sources.}
Assuming a Jansky-to-Kelvin (S/T$_A$) conversion factor of 30 \footnote{ http:$//$docs.jach.hawaii.edu$/$JCMT$/$HET$/$GUIDE$/$het\_guide} %25
 for the JCMT and 41 for APEX\footnote{ http:$//$www.apex-telescope.org$/$telescope$/$efficiency}, and presuming that the total linewidths are respectively of 2 and 1\,km\,s$^{-1}$ for DM\,Tau and TW\,Hya, we derived 1 $\sigma$ noise level on the integrated line  of  0.32\,Jy\,km\,s$^{-1}$ for DM\,Tau and 0.33\,Jy\,km\,s$^{-1}$ for TW\,Hya. We converted to flux density as the sources are unresolved.

Previous observations by \citet{Ceccarelli_etal2004},\citet{Thi_etal2004}, and \citet{Qi_etal2008} give a 1\,$\sigma$ rms of 1.2\,Jy\,km\,s$^{-1}$ for TW\,Hya and 0.88 \,Jy\,km\,s$^{-1}$ for DM\,Tau (assuming a S/T$_{MB}$ conversion factor of 55\,Jy/K for the CSO telescope). 
So, our limits on the integrated intensities are respectively 3.3 and 3.6 times better for DM\,Tau and TW\,Hya.%%\\

%\section{Observed upper limits on \hhd\,column densities}
\section{Results}
\label{sec:res}
To derive upper limits on the \ohhd\,column density from the non-detections, taking into account the disk structure, we used DISKFIT, a radiative transfer code optimized for disks \citep{Pietu_etal2007,Pavlyuchenkov_etal2007}. This code fits a parametric model to the data and the best model is determined by a minimization process. 
Disks can be reasonably described by a parametric  model where all primary quantities are power laws {functions} of the radius \citep[see][]{Dutrey_etal1994,Pietu_etal2007}: surface density $\Sigma(r) = \Sigma_0(r/R_0)^{-p}$, temperature $T(r) = T_0(r/R_0)^{-q}$, velocity $v(r) = V_0(r/R_0)^{-v}$ and scale height $H(r) = H_0(r/R_0)^{-h}$ that controls the vertical density structure. 
The temperature is a simple power law of the radius, no vertical gradient is taken into account. %  $T(r) = T_0(r/R_0)^{-q}$.  
In the peculiar case of non detection, all parameters except the value of $\Sigma_0$ are fixed.  
The surface density of \ohhd\,at the reference radius (noted $\Sigma_0$) is then derived under the LTE hypothesis \citep[valid for densities $>10^5$cm$^{-3}$ see][ Fig.5\,a]{Parise_etal2011}. This parametric approach limits the number of assumptions on the disk structure. 
Table\,\ref{tab:obs} presents the resulting \ohhd\,column densities distributions for several models.

For the two sources, parameters such as position, inclination, and position angle are fixed according to previous CO observations \citep{Pietu_etal2007, Hughes_etal2008}. Two values are adopted for the exponent of the column density distribution $p=0$  (i.e. flat distribution) and $p=-1$ (i.e. surface density linearly increasing with radius) to match the results of chemical modeling \citep[see][and this work]{Willacy_2007}. 

Given the uncertainties on the disks thermal structure, we derived the \ohhd\,column density under two cases for each source, with low (case L) and high (case H) temperature. 
For DM\,Tau, a useful upper limit can be derived from the \dco\ measurement with $T_{100}=30$K and $q=0.6$. This is an upper limit as the \dco\ emission is optically thick enough to trace mainly the disk uppers layers. Less abundant isotopologues, which sample deeper in the disk, as well as other molecules \citep[HCO$^+$, C$_2$H, CN and HCN, see][]{Pietu_etal2007, Henning_etal2010, Chapillon_etal2011} give lower values ($T_{100} \sim $10 -- 15, $q\sim $0 -- 0.4). The law $15\times (r/100)^{-0.4}$ is adopted for the cold case. 
For TW\,Hya, the two temperature cases considered here ($T_{100}=40\mathrm{K}, q=0.2$ and $T_{100}=30\mathrm{K}, q=0.5$) correspond to the two temperature solutions of \citet{Hughes_etal2008}.

The results are also dependent on the external radius. In the $p=0$ case, $R_{out}$ is chosen  in agreement with CO observations (i.e. 720\,AU for DM\,Tau and 200\,AU for TW\,Hya). When the \ohhd\,column density increases with the radius ($p=-1$), the flux is dominated by the outer disk, we adopted a radius of 550\,AU for DM\,Tau in that case. This assumption is in accordance with the ``tapered edge'' model (see Discussion).

\begin{table}
\caption{\ohhd column densities}
\centering
%\begin{center}
\small
\begin{tabular}{lccc|c}
\hline \hline
\multicolumn{4}{c|}{fixed parameters}& 3$\sigma$ upper limit \\
T$_{100}$(K) & q &  R$_{out}$(AU)& p & $\Sigma_{100}$ (cm$^{-2}$) \\ 
\hline
\multicolumn{5}{c}{DM\,Tau}\\
 \,15 & 0.4  & 720 &  0 &$ 1.9\times 10^{12}$ \\
 \,15 & 0.4  & 550 &  -1 &$ 8.0\times 10^{11}$ \\
 \,30 & 0.6  & 720 &  0 &$1.1\times 10^{12}$ \\
 \,30 & 0.6  & 550 &  -1 &$4.5\times 10^{11}$ \\
% \,[10] & [0] & 0?& 720? & 0? &  $4.5 \pm\ 0.9~10^{12}$ & $^b$\\ published by CC04
 \,10 & 0.0 &  720 & 0 & $ 1.3\times 10^{12}$  \tablefootmark{(a)}\\ 
%\hline
%\hline
\hline
\multicolumn{5}{c}{TW\,Hya}\\
%\,20 & 0.4 & 200 & 0 & $1.7\times 10^{12}$  \\
%\,20 & 0.4 & 200 & -1 & $1.3\times 10^{12}$  \\
\,30 & 0.5 & 200 & 0 & $1.4\times 10^{12}$  \\
\,30 & 0.5 & 200 & -1 & $1.0\times 10^{12}$  \\
\,40 & 0.2 & 200 & 0 & $1.3\times 10^{12}$  \\
\,40 & 0.2 & 200 & -1 & $9.0\times 10^{11}$  \\
\hline
\end{tabular}
\normalsize
%\end{center}
\label{tab:obs}
\tablefoot{
Derived \ohhd\,column densities. 
\tablefoottext{a}{Parameters similar to those of \citet{Ceccarelli_etal2004}.} }
\end{table}

% result 
The derived $3\,\sigma$ upper limits on the \ohhd\,column density at 100\,AU is about $10^{12}$cm$^{-2}$ (see Table\,\ref{tab:obs}
). That is $\sim 4$ times lower than the previous estimation of $4.5 \pm 0.9 \times 10^{12}$ cm$^{-2}$ by \citet{Ceccarelli_etal2004} on DM\,Tau. To make a more sensible comparison, the value of $\Sigma_{100}$ was also derived adopting the same disk parameters as \citet{Ceccarelli_etal2004} ($T_{100}=10$\,K, q=0 and p=0). In that case we find $\Sigma_{100}=1.3 \times 10^{12}$ cm$^{-2}$, so a factor 3 lower.

 \citet{Qi_etal2008} have derived a $3\sigma$ upper limit of $5.1 \times 10^{12}$ cm$^{-2}$ on TW\,Hya. Our upper limit is a factor $\sim 5$ lower but because their adopted thermal structure is not given, we cannot fit the same disk model to our data.

\section{Discussion}
\label{sec:dis}

\begin{table*}
\caption{Disk structure adopted for the chemical modeling}%{Chemical parameters}
\label{table:chempar}
\centering
\small
\begin{tabular}{l|cccc|cccc|cc|cc}
\hline
\hline
 & \multicolumn{8}{|c|}{\hh density} & \multicolumn{4}{|c}{Temperature} \\
 & \multicolumn{4}{|c|}{Model 1} & \multicolumn{4}{|c|}{Model 2} & \multicolumn{2}{|c|}{Case L} & \multicolumn{2}{|c}{Case H}\\
 & $\Sigma^{\hh}_0$ (cm$^{-2}$) & $p$ & $R_0$ (AU)& $R_{out}$ (AU)&$\Sigma^{\hh}_0$ (cm$^{-2}$)& $p$ & $R_0$ (AU)&$R_c$ (AU) & $T_{100}$ (K)& $q$ & $T_{100}$ (K)& $q$ \\
\hline
DM\,Tau & $8.4 \times10^{23}$ & 0.85 & 45 & 700 & $9.6 \times 10^{23} $ &0.45 & 45 & 180 & 15  & 0.4 & 30 & 0.63\\
TW\,Hya & $4.3 \times 10^{23}$ & 1.0 & 45 & 200 & $3.3 \times 10^{23}$ & 0.7 & 45 & 30 & 30 & 0.5 & 40 & 0.2\\
\end{tabular}
\normalsize
\end{table*}

Our deep integrations show that \ohhd\,emission, if any, is actually much fainter than previously thought. In this section, we discuss the implications of our non-detections on the amount of CO remaining in gas-phase in the disks mid-plane.

To derive a lower limit on the CO abundance in the disk midplane, we make use of the time-dependent chemical code dedicated to deuterium chemistry presented in Parise et al. (2011) and calculate the \ohhd\,abundance on a grid of densities and temperatures, over several hypothesis on the local CO abundance, grain growth and ionization rate. Using a realistic density and temperature profile for the disk, we then compute the resulting column density of \ohhd\,, and compare it to the one derived from observations. We also derive the emission lines profiles for each of those disk chemical models using the radiative transfer code DISKFIT under LTE.
%To derive a lower limit on CO abundance in the disk mid-plane, we {set} the \hh\,density and temperature disk structures and calculated the \ohhd\,abundance in each point of the disk over several hypothesis on the local CO abundance, grain growth and ionization rate using the time-dependent chemical code dedicated to deuterium chemistry presented in \citet{Parise_etal2011}. The resulting \hhd\,column densities are compared to the one derived from observations. We also derive the emission lines profiles for each of those disk chemical models using the radiative transfer code DISKFIT under LTE.\\

%\subsection{Chemical modeling}
%We ran some chemical models using a code developed by \citet{Du_parise2011}. 
The chemical model was first developed to describe the deuterium chemistry at work in cold prestellar cores \citep{Parise_etal2011}. 
This code takes into account all different spin states of the different molecules and ions. In particular, ortho and para \hh, H$_3^+$ and isotopologues are considered separately, with up-to-date reaction rates from \citet{Hugo_etal2009} and recombination rates for all ions of \citet{Pagani_etal2009}. A simple chemistry of CO and N$_2$ is also included. The reactions are all in gas-phase except the formation of H$_2$, HD and D$_2$ {on} grains. The code is time-dependent, and the developement of deuterium fractionation was shown to be limited by the ortho-\hh\,to para-\hh\,conversion timescale \citep{Pagani_etal2009}. In the mid-plane of protoplanetary disks, we
expect that this conversion has already efficiently happened, and therefore use the steady-state abundances reached at long times.

We obtained the \ohhd\,abundance relative to H$_2$ as a function of density and temperature. We ran the code for several values of the CO abundance (x[CO] = $0, 10^{-6}, 10^{-5}, 10^{-4}$). 
%As the efficiency of the grain chemistry is controlled by the effective grain size, we also tested several value (a = $0.1, 1$ and $10 \mu$m). 
We also varied the effective grain size (a = $0.1, 1$ and $10 \mu$m). 
The abundance of \hhd\,should also depend on the ionization, so we varied the cosmic rays ionization rate ($\xi = 1 \times 10^{-17},3 \times 10^{-17}$ and $1 \times 10^{-16}$s$^{-1}$). The UV field is neglected, which should not be a problem as \hhd\,will in any case be predominantly in the dense regions.%%\\

To estimate the \ohhd\,column densities resulting from the chemical simulations we need a model of the disk \hh\,density and temperature. %We still adopted a model where the parametric one. 
For the surface density we adopted two different models:
\begin{itemize}
\item the simple model where the \hh\,surface density is described by a power law  $\Sigma^{H_2}(r) = \Sigma^{H_2}_0(r/R_0)^{-p}$, truncated at a given radius ($R_{out}$), hereafter Model 1,
\item and a model where the \hh\,surface density is an exponentially tapered power law as would result from viscous spreading of the disk 
$\Sigma^{H_2}(r) = \Sigma^{H_2}_0(r/R_0)^{-p} \mathrm{exp}(-(r/R_c)^{(2-p)})$ 
hereafter Model 2 \citep[see also][]{Hughes_etal2008}.
\end{itemize}
We adopted the same temperature law as previously, {without} vertical gradient. Parameters are summarized in Table\,\ref{table:chempar}. To derive local densities, we further assumed hydrostatic equilibrium.

\begin{figure}%[h!]
\centering
\includegraphics[width=7.5cm]{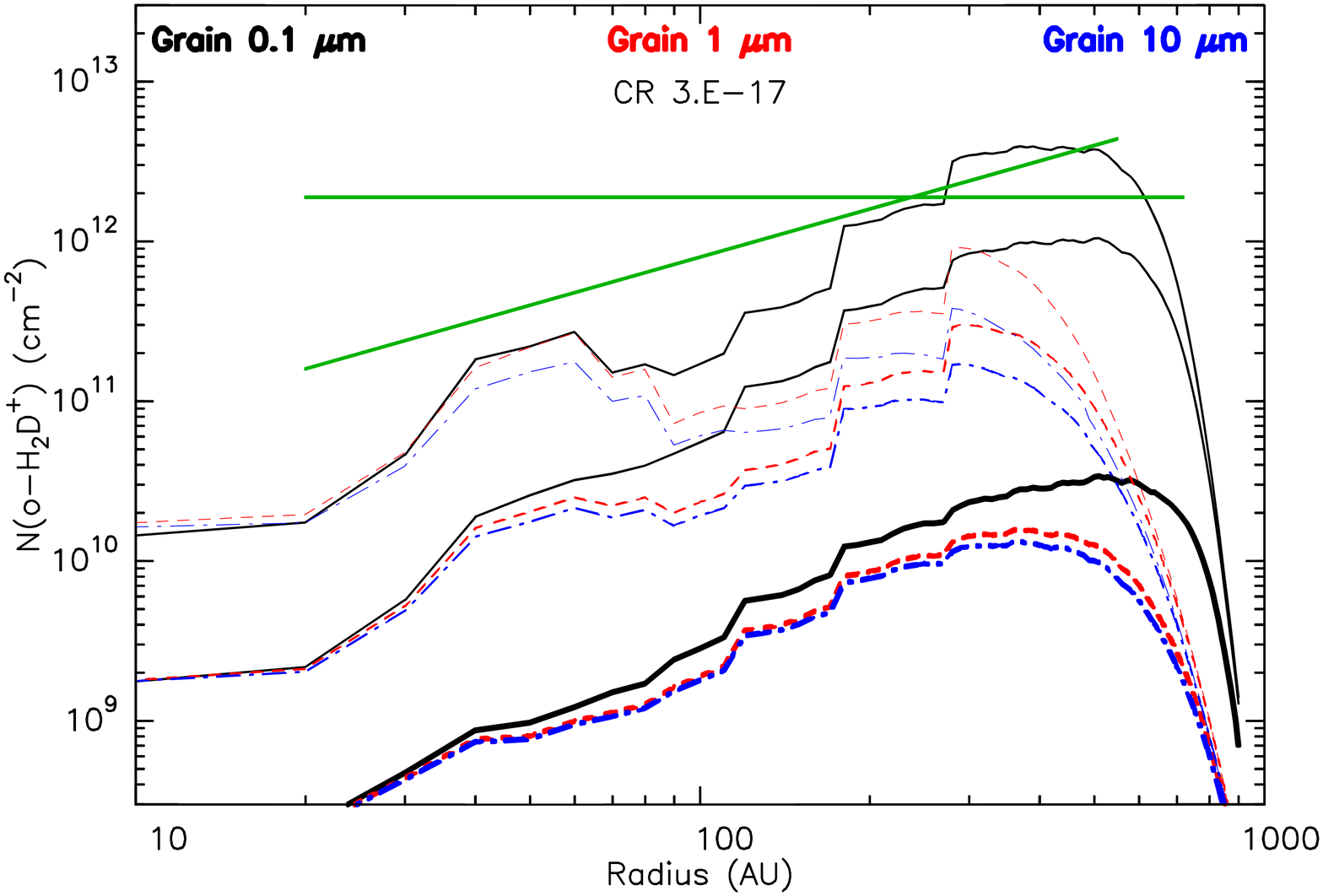}\\%{newfig5.eps}\\%
\includegraphics[width=7.5cm]{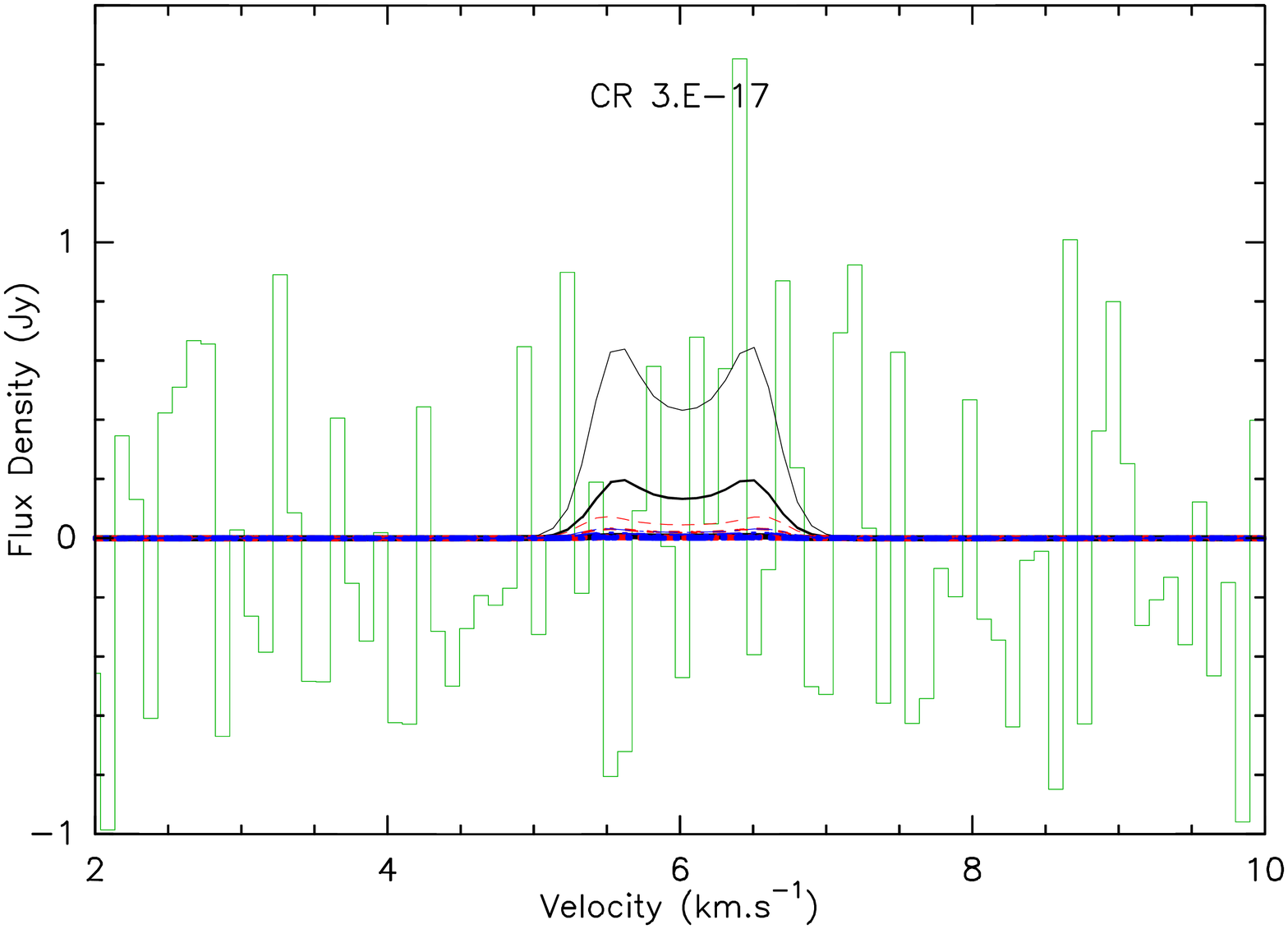}%{modif3.eps}%
\caption{Results from the chemical modeling for DM\,Tau, case L, Model 2 and $\xi=3\times 10^{-17}$s$^{-1}$. Top: Predicted \ohhd\,column densities as a function of radius for several values of grain size:  a=0.1\,$\mu$m (black), 1 (red) and 10 (blue), and CO abundance: x[CO]= $10^{-6}$ (thin lines) $10^{-5}$ (medium) and $10^{-4}$ (thick). The green lines correspond to the best $3\sigma$ upper limits derived from observations with p=0 and p=-1. Bottom: Observed spectrum (green) and predicted line profiles (same color and thickness convention). %\textbf{ A:\,x[CO]=$10^{-6}$, a=0.1\,$\mu$m, B:\,x[CO]=$10^{-5}$, a=0.1\,$\mu$m.}
}
\label{fig:mod}
\end{figure}

\begin{figure}
\label{fig:dep}
%\resizebox{\hsize}{!}{\includegraphics[angle=270]{plot-cr-gr-xco.eps}}
\begin{center}
%\centering
\includegraphics[angle=270,width=8.5cm]{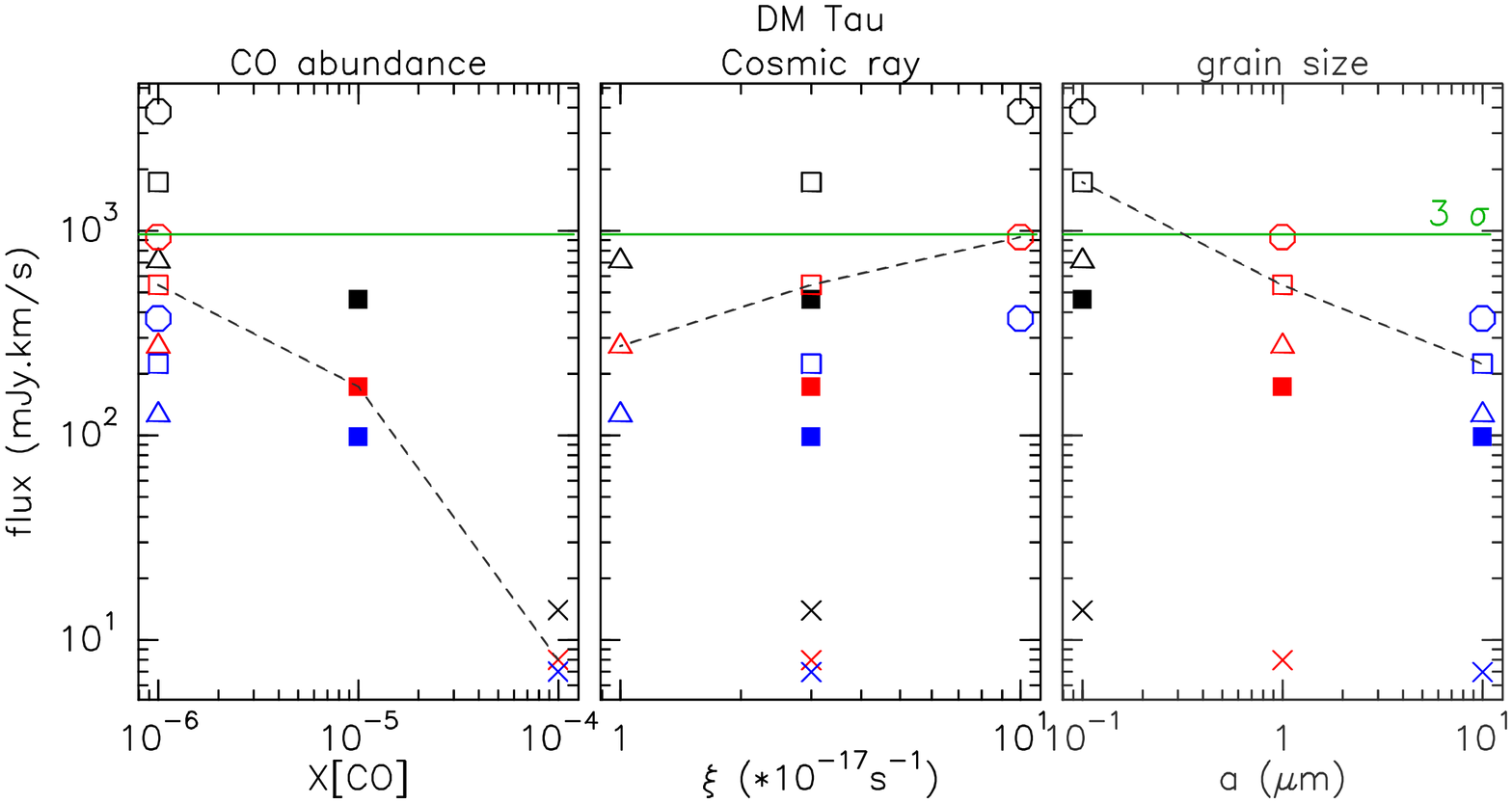}
\end{center}
\caption{\ohhd\,predicted flux as a function of CO abundance (left), cosmic ray (middle) and grain size (right) for DM\,Tau.}
Grain size: a=0.1\,$\mu$m (black), 1 (red) and 10 (blue). 
Cosmic ray ionization rate: $\xi=1\times10^{-17}$ (triangle), $3\times10^{-17}$ (square) and $1\times10^{-16}$\,s$^{-1}$ (octagon). 
CO abundance: x[CO]= $10^{-6}$ (open marker) $10^{-5}$ (filled marker) and $10^{-4}$ (cross).
The green line corresponds to our $3\sigma$ limit derived from observations. 
The dashed lines illustrate the dependencies.
\end{figure}

\begin{table}
 \caption{Predicted \ohhd\,line fluxes (mJy\,km\,s$^{-1}$) for DM\,Tau}
 \label{tab:matrix-cr}
\small
 \begin{tabular}{cc}
   \ohhd\,fluxes for $\xi = 3 \times 10^{-17}$s$^{-1}$ 
 &  \ohhd\,fluxes for x[CO] = $10^{-6}$\\
   \begin{tabular}{c|ccc}
     \hline \hline 
     {x[CO]}   & \multicolumn{3}{c}{a $(\mu$m)}\\
      &  0.1 & 1 & 10\\
         \hline %\cline{3-5}
     %\multirow{3}{0.1cm}{\rotatebox{90}{x[CO]}} & $10^{-4}$ & 14 & 8 & 7 \\
       $10^{-4}$ & 14 & 8 & 7 \\
       $10^{-5}$ & 465 & 174 & 99 \\
         $10^{-6}$ & 1730 & 544 & 224\\   
     \hline
   \end{tabular}
   &
%   \ohhd\,fluxes for x[CO] = $10^{-6}$\\
   \begin{tabular}{c|ccc}
     \hline \hline 
     {$\xi$  (s$^{-1}$)} & \multicolumn{3}{c}{a $(\mu$m)}\\
     %\multirow{3}{0.1cm}{\rotatebox{90}{$\xi ( 10^{-17}$s$^{-1})$}}  & & 0.1 & 1 & 10\\
       $\times 10^{-17}$ & 0.1 & 1 & 10\\
     \hline %\cline{3-5}
       10  & 3830 & 932 & 373 \\
       3   & 1730 & 544 & 224 \\
       10  & 713 & 273 & 126\\   
     \hline
   \end{tabular}%%\\
\normalsize
 \end{tabular}
\end{table}

For DM\,Tau, the disk parameters mainly come from the analysis performed by \citet{Pietu_etal2007}. The density models are derived from the continuum observations of \citet{Guilloteau_etal2011}. Model 1 is arbitrarily extended up to an external radius of 700\,AU (in agreement with CO emission). 

For TW\,Hya, the parameters are derived from \citet{Hughes_etal2008}. Note that their ``cold'' case applies to the density Model 1 and the ``hot'' case to the Model 2.

The disks are expected to have vertical temperature gradients which we ignore in this simple modeling. As a consequence, the abundance of \ohhd\ above the mid-plane is underestimated \citep[see for instance][Fig.\,1]{Ascencio_etal2007}. Nevertheless, as the \hh\ density above the mid-plane is significantly lower, the contribution of this additional layer to the total \ohhd\ column density is negligible \citep[see Fig. 6 and 7 of][]{Willacy_2007}. For DM\,Tau the two adopted temperature profiles represent two extreme cases, one only considering the low mid-plane temperature, the other using the higher values sampled by \dco\ at 1-2 scale heights.

The top panel of Fig.\,\ref{fig:mod} presents the distribution of the \ohhd\,column density for DM\,Tau, Model 2 case L and the standard value of the cosmic ray ionization $\xi = 3 \times 10^{-17}$ s$^{-1}$. All models with CO depletion show a region of high column density at R = 40--50\,AU (low temperature case) or R = 150--200\,AU (high temperature case). In each case, this corresponds to a kinetic temperature around 20 K. These models display also a region of low \ohhd\,column density for T$\sim$15K.  This is caused by the rapid evolution of the ortho/para ratio of H$_2$D$^+$ with temperature in this range (the ratio o/p decreases by a factor of 3 between 20K and 15K for a density of 10$^6$\,cm$^{-3}$, and x(CO)=10$^{-6}$). This effect has already been shown in previous works, see \citet[][their Fig.\,6]{Flower_etal2004} and \citet[][their Fig.\,4]{Sipila_etal2010}.
%This might be related to the evolution of the ortho/para ratio with temperature, see \citet[][their Fig. 6]{Flower_etal2004} and \citet[][their Fig.\,4]{Sipila_etal2010}. 
Apart from that feature (which is much more pronounced for x[CO]=0), the \ohhd\,column density increases with the radius in Model 1 \citep[in qualitative agreement with ][]{Willacy_2007}. In Model 2 this increase is tapered off by the exponential fall of the H$_2$ surface density. For DM\,Tau, the \ohhd column density peaks around R = 550\,AU where it reaches $10^{13}$cm $^{-2}$. 
The disk of TW\,Hya being small (R$_{out} \sim$ 200\,AU), the temperature is high in the whole disk, so the \ohhd\,column density is low whatever the CO abundance (column density below $10^{12}$cm$^{-2}$ if we exclude the high ionization case) and the models are not constraining. 
The predicted integrated spectra for each of the models are computed using DISKFIT. The bottom panel of Fig.\,\ref{fig:mod} presents the predicted spectra together with observational data for DM\,Tau, Model 2 case L and the standard value of the cosmic ray ionization. 
The predicted \ohhd\,line fluxes for DM\,Tau are presented in Table \ref{tab:matrix-cr} and Fig.\ref{fig:dep}.  The integrated line flux seems to vary roughly like $\xi^{0.5}$ and a$^{-0.5}$. The dependancy over the CO abundance is more complicated. 

From Fig.\,\ref{fig:dep} one can see that unless the ionization rate is very high, the current upper limits are not very stringent on the disk parameters. At best only the small grains, CO depleted by a factor 100 and $\xi > 3\times 10^{-17}$s$^{-1}$ case are excluded in the DM\,Tau case. With a CO depletion of order 10 or so, any grain size would fit the current data. The chemical models are not constraining for TW\,Hya.

{
Compared to the earlier study of \citet{Ceccarelli_dominick2006}, which did not consider the ortho and para \hhd\ separately, our chemical model including all spin-dependent reaction rates predicts column densities on average 5-10 times weaker.  Accordingly, as the \ohhd\,line emission is mostly optically thin, except for the highest column densities (above a few $10^{12}$\,cm$^{-2}$ ), we predict flux densities 3 to 5 times lower than \citet{Ascencio_etal2007}.
%\citet{Ascencio_etal2007}
%Our model yields typical flux densities lower by about a factor 3-4 from those of \citet{Ascencio_etal2007} \citep[based on ][model]{Ceccarelli_dominick2006}, and are more in agreement with the more stringent upper limits we have obtained.
}

\section{Implications for ALMA}
\label{sec:alma}
In this section, we discuss the feasibility of observational improvements compared to the present work, using the ALMA interferometer. Figure 3 demonstrates that an increase in flux  sensitivity by at the very least a factor of 10 (note that the emission is optically thin)  compared to the present observations is required to secure a convincing detection  of \hhd\,(in the framework of the chemical models of Sect 4). An even higher  sensitivity would be required to disentangle the different models.
%\textbf{ copie mail stephane}
The current upper limits, obtained with single-dishes of 12 - 15 m diameter and integration times 5--10 hours, show that only larger instruments can  expect detecting \hhd. 
ALMA is the most  promising: in single-dish mode, the 50 12-m antennas will bring a gain of 7 in sensitivity compared to APEX (assuming similar receiver performances). However, in interferometric mode, even in its compact configuration,
%ALMA is the most promising, but even in its compact  configuration,
 the angular resolution of the main array will be around 1$''$, so that the sources will be highly resolved out. In such cases, the brightness sensitivity becomes the relevant parameter. In the beam of the current instruments, the observed discs are diluted by about a factor 3 to 4, as the APEX beam size is $14''$ compared to an outer disc diameter around $7-8''$. However, the aperture filling factor of ALMA is only about 40 \%, compared to filled aperture for APEX. Hence, in a similar integration time, and assuming similar noise performance, the ALMA brightness sensitivity should enable to reach a detection level of column densities only 2 times smaller that the current level in each ALMA synthesized beam. Further spatial smoothing would improve the sensitivity to extended structure, but only moderately, as this implies degrading the weight of the longest baselines. Using the ALMA compact array (ACA) may be more appropriate: the collecting area of the 12 7-m antennas is 4 times that of APEX, so accounting for an effective filling factor around 30-40 \% also leads to similar brightness sensitivities  as for the 12-m antennas compact configuration.

While detailed predictions will require proper knowledge of the ALMA performances at the difficult frequency of 372.4\,GHz,  these simple considerations show that \ohhd\,is unlikely to become a major ``workhorse'' to study e.g. the kinematics of disks.

\section{Conclusion}
\label{sec:cc}
\begin{itemize}
\item We have improved the sensitivity to \ohhd\,by a factor 3 over previous observations. No \ohhd\,is detected, in contradiction with previous tentative detection.
\item With our chemical model, TW Hya is in any case too warm or not dense enough to provide detectable \ohhd\,at the current sensitivity.
\item  The limit on DM Tau is incompatible with very high CO depletion \textbf{ (x[CO]$<10^{-6}$)} and small grains \textbf{($0.1\,mu$m)}. However, it is not discriminant with more reasonable values of the grain sizes and/or CO abundances.
\item Much more sensitive observations will be required to obtain significant constraints. 
Even with the powerful ALMA interferometer, it will be extremely difficult to reach a sensitivity that will allow to constrain the ionization level and the amount of CO in the disk mid-planes by means of the observation of \ohhd.
%Only ALMA can provide this. 
\end{itemize}

\begin{acknowledgements}

We acknowledge Anne Dutrey for many fruitful discussions. Remo Tilanus kindly provided us with the JCMT data. %\textbf{ We are gratefull to the APEX staff and in particular S. Leurini and F. Wyrowsky for performing the APEX observations.}\\
E.C.,B.P. and F.D are supported by the {\it Deutsche Forschungsgemeinschaft} (DFG) under the Emmy Noether project PA 1692/1-1.
S.G. acknowledges financial support by the French program PCMI from CNRS/INSU.
\end{acknowledgements}

\bibliography{bib}
\bibliographystyle{aa}
\bibdata{bib}
\bibstyle{aa}

\end{document}